\documentclass{cjaa}                   % preprint, the final version for publication
\usepackage{graphicx}                   %for PS/EPS graphics inclusion, new
\input{epsf.sty}                        %for PS/EPS graphics inclusion, old
\input{psfig.sty}                       %for PS/EPS graphics inclusion, old

\begin{document}

   \title{Fitting Formulae for the Effects of Binary Interactions on Lick Indices and Colours of Stellar Populations
%\footnotetext{$*$ Supported by the National Natural Science Foundation of China.}
}
%   \subtitle{I. Place Your Subtitle Here}

   \volnopage{Vol.0 (200x) No.0, 000--000}      %%preserved for Editor. DOn't remove!
   \setcounter{page}{1}          %%starting page, preserved for Editor. DOn't remove!

   \author{Zhong-Mu Li
      \inst{1}\mailto{}
%% Please move "\mailto{}" to the corresponding author of the paper.
   \and Zhan-Wen Han
      \inst{2}
      }
   \offprints{Z.-M. Li}                   %% is disabled in fact

   \institute{Dali Uinversity, Dali, 671003, China\\
             \email{zhongmu.li@gmail.com}
%% offprint requests will be sent. Note to pair \mailto{} with \email{}
        \and
        National Astronomical Observatories/Yunnan Observatory, the Chinese
Academy of Sciences, Kunming, 650011,
China\\
          }

   \date{Received~~2007 Dec. 17; accepted~~2001~~month day}

   \abstract{
    More than about 50\% stars are in binaries, but
    most stellar population studies take single star stellar
    population (ssSSP) models, which do not take binary interactions into
    account. In fact, the integrated peculiarities of ssSSPs are various from
    those of stellar populations with binary interactions (bsSSPs).
    Therefore, it is necessary to investigate the effects of binary
    interactions on the Lick indices and colours of populations detailedly.
    We show some formulae for calculating the difference between
    the Lick indices and colours of bsSSPs, and those of ssSSPs. Twenty-five Lick
    indices and 12 colours are studied in the work.
    The results can be conveniently used for calculating the effects
    of binary interactions on stellar population studies and for
    adding the effects of binary interactions into present ssSSP
    models. The electrical data and fortran procedures of the paper can be obtained on request to the authors.
   }

   \authorrunning{Z.-M. Li \& Z.-W. Han}            %author_head in even pages
   \titlerunning{Binary Interactions in Stellar Populations}  % title_head in odd pages

   \maketitle
%% The author head (on even pages) and the title head (on odd pages) will be
%% automatically extracted from \author{} and \title{}. Whenever the title is too long,
%% you will be asked to supply a shorter one by inserting either \authorrunning{} or
%% \titlerunning{} before \maketitle. Anyway, you can specify your own heads in advance.
%%
%%
%% Note: In the following text body of your manuscript, please note several differences from
%%       other major journals:
%% (1) \subsection{Please Capitalize the First Letter of Each Notional Word in Subsection Title}
%% (2) Please Capitalize the First Letter of Each Notional Word in table's caption

%
%________________________________________________ sections below
%
\section{Introduction}           %% first-level sections will be auto-capitalized
\label{sect:intro}
%\hspace{15pt}%                   %% preserved for Editor
In the golden era for studying the formation and evolution of
galaxies, evolutionary stellar population synthesis has been an
important technique for such works, as some stellar characteristics
(e.g., stellar age and metallicity) of galaxies can be determined
via this technique. Many stellar population synthesis models, e.g.,
\cite{Worthey:1994}, \cite{Buzzoni:1995}, \cite{Bressan:1994},
\cite{Vazquez:2005}, \cite{Bruzual:2003}, \cite{Fioc:1997},
\cite{Vazdekis:2003}, \cite{Delgado:2005}, and \cite{Zhang:2005hr},
were brought forward and have been widely used for stellar
population studies. However, the above models except the one of
\cite{Zhang:2005hr} are single star stellar population (ssSSP)
models that did not take the effects of binary interactions into
account. According to the results of \cite{Han:2001}, more than 50\%
stars of the Galaxy are in binaries and binaries evolve differently
from single stars. The real stellar populations of galaxies and star
clusters consist of not only single stars, but also binary stars.
Thus ssSSPs are different from the real populations of galaxies and
star clusters. In fact, binary evolution can affect the integrated
peculiarities (e.g., the spectral energy distributions in UV bands)
of stellar populations significantly (see, e.g., \cite{Han:2007}).
Therefore, the effects of binary evolution should be taken into
account when modeling the stellar populations of galaxies and star
clusters.

A few works have been tried to investigate the effects of binary
evolution on stellar population synthesis. For example,
\cite{Zhang:2005hr} tried to model populations via binary stars. In
addition, \cite{Li:2008database} built an isochrone database for
quickly modeling binary star stellar populations (bsSSPs) and a
rapid model (hereafter $RPS$ model) for both ssSSPs and bsSSPs. In
special, \cite{Li:2008how} investigated the detailed effects of
binary interactions on the results of stellar population synthesis
and the results of stellar population studies. The results can help
us to understand how the results obtained via ssSSPs are different
from those obtained via bsSSPs, when taking H$\beta$--[MgFe]
(\cite{Thomas:2003}) and two-colour methods. According to the
results of \cite{Li:2008how}, when we use ssSSP models to measure
the stellar ages and metallicities of galaxies, we will obtain
obviously younger ages or lower metallicities compared to the real
values of populations, using H$\beta$--[MgFe] and two-colour
methods, respectively. However, there is no clear relation between
the real metallicities and fitted (via ssSSPs) results of
populations. One please refer to \cite{Li:2008how} for more details.
In this case, it is difficult to get more accurate information about
the stellar metallicities of galaxies via ssSSP models, and then the
chemical evolution of galaxies. Furthermore, the previous work only
shows the results for H$\beta$--[MgFe] method, when taking Lick
indices for works, but some other methods and indices are also used
in investigations. Thus it is necessary to investigate the effects
of binary interactions on the results of stellar population studies
obtained via various Lick indices further. The metallicity range of
above bsSSP models (\cite{Zhang:2005hr}, \cite{Li:2008database})
seems not wide enough (see \cite{Li:2006}), as it only covers the
metallicity range poorer than 0.03 ($Z \le$ 0.03). This is limited
by the star evolution code. If we can give the relation between the
effects of binary interactions and the stellar-population parameters
(age and metallicity), we will be able to understand the populations
of galaxies and star clusters further, and more detailed
investigations about galaxy formation and evolution will have in the
future. Therefore, it is valuable to study how the effects of binary
interactions on integrated peculiarities of populations change with
stellar age and metallicity. We have a try in this work. As a
result, a few formulae for describing the relations between the
effects of binary interactions on integrated indices (Lick indices
and colours indices) and stellar-population parameters are
presented.

The structure of the paper is as follows. In Sect. 2 we introduce
the stellar population model used in the paper. In Sect. 3 we show
the fitting formulae for the changes of 25 Lick indices caused by
binary interactions when comparing to those of ssSSPs. In Sect. 4 we
give similar investigations to 12 colours of populations. Finally,
we give our discussion and conclusion in Sect. 5.

%% ChJAA editors DID NOT use \cite{} for citation, \ref and \label for
%% cross-references of Table/Figure in publication version.
%% ChJAA editors prefered you giving a citation as 'Michel et al. 1992', and
%% writting Table~1 or Fig.~1 and so forth. However, that will make authors
%% inconvenient in adjusting/adding/removing text, tables or figures. Anyway,
%% authors can use \cite, \citep and \citet as widely used in other journals.
%% ChJAA editors are moving to use a more flexible LaTeX source.

\section{Stellar population model used in the paper}
\label{sect:Obs}
%\hspace{15pt}%                   %% preserved for Editor
The $RPS$ model of \cite{Li:2008database} is used in this
investigation, because there is no more suitable model. The model
calculated the integrated peculiarities (0.3 $\rm \AA$ SEDs, Lick
indices and colours) of both bsSSPs and ssSSPs with two widely used
initial mass functions (IMFs) (Salpeter and Chabrier IMFs). Each
bsSSP contains about 50\% stars that are in binaries with orbital
periods less than 100\,yr (the typical value of the Galaxy, see
\cite{Han:1995}). Binary interactions such as mass transfer, mass
accretion, common-envelope evolution, collisions, supernova kicks,
angular momentum loss mechanism, and tidal interactions are
considered when evolving binaries via the rapid stellar evolution
code of \cite{Hurley:2002}. Therefore, the $RPS$ model is suitable
for studying the effects of binary interactions on stellar
population synthesis. The details about the model can be seen in the
paper of \cite{Li:2008database} and \cite{Li:2008how}. For
convenience, we take stellar populations with Salpeter IMF for our
standard investigations in the work, but the results obtained via
populations with Chabrier IMF are also presented.

\section{Fitting formulae for the effects of binary interactions on 25 Lick indices}
\label{sect:data}
%\hspace{15pt}%                   %% preserved for Editor
Lick indices are the most widely used indices in stellar population
studies, because they can disentangle the well-known stellar
age--metallicity degeneracy (\cite{Worthey:1994}). Making use of an
age-sensitive index (e.g., H$\beta$) together with a
metallicity-sensitive index (e.g., [MgFe], see \cite{Thomas:2003}),
the stellar age and metallicity of a population can be determined.
Thus to investigate the effects of binary interactions on the Lick
indices of stellar populations is important. The work of
\cite{Li:2008how} showed that binary interactions make the H$\beta$
index less while some metal-line indices larger compared to those of
ssSSPs. It leads to younger age estimate when we take ssSSPs for
works. However, in that work, only the results obtained via
H$\beta$--[MgFe] method are compared to the real values of
populations. Some other Lick indices, e.g., Mg2, H$\delta_{\rm A}$,
and H$\gamma_{\rm A}$, are also widely used in studies (e.g.,
\cite{Gallazzi:2005}). Therefore, it is necessary to study the
effects of binary interactions on more Lick indices and give the
quantitative relations between binary effects and stellar-population
parameters. Here we study on 25 widely used indices and fit the
relations between the changes caused by binary interactions and the
stellar-population parameters (age and metallicity), via a
polynomial fitting method. The results can be used to calculate the
differences between the 25 Lick indices of two kinds of populations
with small errors (typically less than 0.03 ${\rm \AA}$ or mag). All
Lick indices are on the Lick system (see, e.g.,
\cite{Worthey:1994licksystem}). When comparing to ssSSPs, the
effects of binary interactions on Lick indices can be calculated
from stellar age and metallicity, by
\begin{equation}
    \Delta I = \sum_{i=1}^{5}
({\rm C}_{i{\rm 1}} + {\rm C}_{i{\rm 2}}Z + {\rm C}_{i{\rm 3}}Z^{\rm
2})t^{i-1},
\end{equation}
where $\Delta I$ is the change of a Lick index caused by binary
interactions, and $Z$ is stellar metallicity, while $t$ is stellar
age. The detailed coefficients for our standard investigation are
shown in Tables 1 and 2. Those for populations with Chabrier IMF are
shown in the Appendix. For clearly, in Figs. 1, 2, and 3, we compare
the changes calculated by equation (1) with the original values
obtained in the work. Note that we only show the fittings for 12
widely used Lick indices here, because the fittings for other
indices are similar. As we see, for the indices shown, the values
calculated by the above equation are consistent with those obtained
directly by comparing the Lick indices of bsSSPs and ssSSPs, with
typical errors of 0.03\,${\rm \AA}$ or mag. Therefore, the fitting
formulae presented can be used to calculate the differences of Lick
indices of bsSSPs and ssSSPs, using the age and metallicity of
populations. In addition, the results show that binary interactions
make age-sensitive indices (e.g., H$\beta$, $H\delta_{\rm A}$,
$H\delta_{\rm F}$, $H\gamma_{\rm A}$, $H\gamma_{\rm F}$) of a bsSSP
larger than those of its corresponding (with the same age and
metallicity) ssSSP, while making metallicity-sensitive indices
(e.g., Mg or Fe indices) of a bsSSP less than that of its
corresponding ssSSP. This is similar to that shown in the paper of
\cite{Li:2008how}. Furthermore, it is shown that the differences
between Lick indices of bsSSPs and ssSSPs increase with age when
stellar age is small ($<$ about 2.5\,Gyr), and they decrease with
age for larger age. This results from the star sample (i.e., the
fraction of binaries and the relation between the masses of the two
components of each binary) of stellar populations. As a whole, the
values calculated via the fitting formulae obtained by the paper
reproduce the evolution of the difference between Lick indices of
bsSSPs and ssSSPs.
\begin{table}[h]
\caption[]{Coefficients for equation (1). The coefficients are
obtained via stellar populations with Salpeter IMF and can be used
for populations younger than 4\,Gyr (Age $<$ 4\,Gyr).} \label{Tab:4}
\begin{center}\begin{tabular}{lcrrrrr}
\hline\hline\noalign{\smallskip}%\scriptsize
Index&$j$ &${\rm C_{1j}}$ &${\rm C_{2j}}$ &${\rm C_{3j}}$ &${\rm C_{4j}}$ &${\rm C_{5j}}$\\
\hline
                  &1 &    0.0061004 &   -0.0124136 &    0.0044131 &   -0.0010174 &    0.0000862 \\
CN$_{\rm 1}$      &2 &   -0.7411409 &    3.6764976 &   -3.4540761 &    0.9481077 &   -0.0789850 \\
                  &3 &   30.8122734 & -119.0088796 &   98.2842545 &  -25.5632396 &    2.0732871 \\
\hline                                                                                          \\
                  &1 &    0.0031344 &   -0.0060648 &    0.0018661 &   -0.0005851 &    0.0000607 \\
CN$_{\rm 2}$      &2 &   -0.3472402 &    2.6820361 &   -2.8891002 &    0.8270163 &   -0.0703722 \\
                  &3 &   20.0763558 &  -92.9530724 &   84.0072412 &  -22.6637506 &    1.8767287 \\
\hline                                                                                          \\
                  &1 &    0.0116927 &   -0.0652076 &    0.0542984 &   -0.0142708 &    0.0011505 \\
Ca4227            &2 &    2.6086437 &    1.4449921 &   -5.0762513 &    1.6986709 &   -0.1529064 \\
                  &3 &  -71.6010754 &   65.3934655 &   -2.6926390 &   -6.3483260 &    0.8599647 \\
\hline                                                                                          \\
                  &1 &    0.5779073 &   -1.3565672 &    0.6261188 &   -0.1240817 &    0.0083961 \\
G4300             &2 &  -50.8998137 &  134.5958812 &  -94.7989188 &   23.2438808 &   -1.8011864 \\
                  &3 & 1740.9893518 &-4249.3881941 & 2828.2830757 & -668.7058103 &   50.9809128 \\
\hline                                                                                          \\
                  &1 &    0.1013725 &   -0.1526486 &   -0.0164836 &    0.0124241 &   -0.0013244 \\
Fe4383            &2 &   11.0894652 &   -9.0004221 &  -12.0126956 &    4.9108205 &   -0.4569935 \\
                  &3 &  -75.5186140 & -252.7173536 &  549.1707675 & -171.1008580 &   14.7310943 \\
\hline                                                                                          \\
                  &1 &   -0.0047432 &    0.0186972 &   -0.0197387 &    0.0040785 &   -0.0002709 \\
Ca4455            &2 &    6.5346797 &  -11.0259844 &    2.2737411 &    0.1733499 &   -0.0467838 \\
                  &3 &  -98.3181765 &   98.4157661 &   49.1921063 &  -27.0189711 &    2.8005540 \\
\hline                                                                                          \\
                  &1 &    0.0550772 &   -0.1274444 &    0.0222163 &   -0.0009561 &   -0.0000671 \\
Fe4531            &2 &   18.2809537 &  -32.2387726 &   11.1472063 &   -1.0492667 &   -0.0001559 \\
                  &3 & -313.2684866 &  374.3927838 &  -24.2302367 &  -31.1448033 &    4.4257825 \\
\hline                                                                                          \\
                  &1 &   -0.0717857 &    0.2203558 &   -0.1810129 &    0.0435325 &   -0.0033410 \\
Fe4668            &2 &   27.2930010 &  -66.0581381 &   35.0658395 &   -6.9644712 &    0.4740158 \\
                  &3 & -444.6620029 &  913.3400658 & -347.7358123 &   51.9244201 &   -2.7150328 \\
\hline                                                                                          \\
                  &1 &   -0.4120752 &    0.9250669 &   -0.3741119 &    0.0622541 &   -0.0036591 \\
H$_\beta$         &2 &   26.4414835 &  -74.8894421 &   48.1881455 &  -11.2855830 &    0.8626001 \\
                  &3 & -805.0295817 & 2280.7328848 &-1494.4764420 &  349.3195439 &  -26.6678020 \\
\hline                                                                                          \\
                  &1 &   -0.0313311 &   -0.0500534 &   -0.0035580 &   -0.0037280 &    0.0007775 \\
Fe5015            &2 &   61.9785871 & -107.1406403 &   40.6549692 &   -4.5784440 &    0.0744308 \\
                  &3 &-1490.9225648 & 2286.4644788 & -734.9273433 &   46.6787692 &    3.5030950 \\
\hline                                                                                          \\
                  &1 &   -0.0011669 &    0.0020035 &   -0.0025450 &    0.0006800 &   -0.0000553 \\
Mg$_{\rm 1}$      &2 &    1.0977834 &   -1.6954874 &    0.4725610 &   -0.0265407 &   -0.0021062 \\
                  &3 &  -23.8411177 &   30.5518215 &   -4.5747311 &   -1.0811535 &    0.1935160 \\
\hline                                                                                          \\
                  &1 &   -0.0036972 &    0.0016391 &   -0.0002221 &   -0.0002245 &    0.0000340 \\
Mg$_{\rm 2}$      &2 &    2.6546194 &   -4.0910074 &    1.1294327 &   -0.0429887 &   -0.0081361 \\
                  &3 &  -63.2072524 &   93.1999501 &  -25.0701341 &    0.5006265 &    0.2481080 \\
\hline                                                                                          \\
                  &1 &   -0.1159519 &    0.0564072 &    0.0367218 &   -0.0207147 &    0.0023015 \\
Mg$_{\rm b}$      &2 &   36.6456419 &  -47.5077586 &    8.8070472 &    1.2388375 &   -0.2750762 \\
                  &3 & -857.5383388 & 1080.0808279 & -207.5498638 &  -26.8404813 &    6.2521890 \\
\hline                                                                                          \\
                  &1 &    0.1443967 &   -0.2658498 &    0.0907173 &   -0.0111382 &    0.0003331 \\
Fe5270            &2 &   -8.5450001 &   12.6759249 &   -7.5470939 &    1.5641500 &   -0.0991560 \\
                  &3 &  306.8420600 & -637.6523032 &  403.3286320 &  -89.3943857 &    6.3626512 \\
\noalign{\smallskip}

\noalign{\smallskip}\hline
\end{tabular}\end{center}
\end{table}

\addtocounter{table}{-1}
\begin{table}[b]
\centering \caption[]{--continued.} \label{Tab:4}
\begin{center}\begin{tabular}{lcrrrrr}
\hline\hline\noalign{\smallskip}%\scriptsize
Index&$j$ &${\rm C_{1j}}$ &${\rm C_{2j}}$ &${\rm C_{3j}}$ &${\rm C_{4j}}$ &${\rm C_{5j}}$\\
\hline                                                                                        \\
                  &1 &    0.1121199 &   -0.2143513 &    0.0776458 &   -0.0095392 &    0.0002385 \\
Fe5335            &2 &   -6.1989089 &   11.7530508 &   -8.2144476 &    1.6722406 &   -0.0990246 \\
                  &3 &  144.7440581 & -352.4412737 &  218.9898664 &  -44.2594619 &    2.7694391 \\
\hline                                                                                          \\
                  &1 &   -0.0167597 &   -0.0094937 &   -0.0069105 &    0.0022746 &   -0.0001900 \\
Fe5406            &2 &   18.1518764 &  -28.2532163 &    9.0539150 &   -0.8243859 &   -0.0014094 \\
                  &3 & -423.9701730 &  596.6530611 & -158.4751506 &    4.4556600 &    1.3344343 \\
\hline                                                                                          \\
                  &1 &    0.0643501 &   -0.1189715 &    0.0352640 &   -0.0029869 &   -0.0000373 \\
Fe5709            &2 &   -6.8927462 &    9.4915829 &   -2.5896330 &    0.0686243 &    0.0256903 \\
                  &3 &  206.1312747 & -349.9572257 &  151.0604011 &  -22.2154935 &    0.9194484 \\
\hline                                                                                          \\
                  &1 &   -0.0193549 &    0.0201941 &   -0.0227794 &    0.0068166 &   -0.0006059 \\
Fe5782            &2 &    5.8734339 &   -8.3105593 &    3.3800221 &   -0.6333537 &    0.0435585 \\
                  &3 & -145.6309054 &  234.0700497 & -130.6085667 &   28.1896236 &   -2.0267932 \\
\hline                                                                                          \\
                  &1 &   -0.1041283 &    0.0782159 &   -0.0131910 &   -0.0028878 &    0.0005590 \\
Na$_{\rm D}$      &2 &   35.8174765 &  -52.1821670 &   16.3021886 &   -1.2441912 &   -0.0404260 \\
                  &3 & -928.3347064 & 1387.5693179 & -484.7687597 &   49.3839423 &   -0.3018862 \\
\hline                                                                                          \\
                  &1 &   -0.0046972 &   -0.0031675 &    0.0086166 &   -0.0031273 &    0.0003031 \\
TiO$_{\rm 1}$     &2 &    2.5376502 &   -3.1526728 &    0.3308384 &    0.2054016 &   -0.0316500 \\
                  &3 &  -70.3829758 &   95.4015845 &  -17.9727317 &   -3.0711691 &    0.6438944 \\
\hline                                                                                          \\
                  &1 &   -0.0089825 &   -0.0020920 &    0.0096106 &   -0.0036755 &    0.0003656 \\
TiO$_{\rm 2}$     &2 &    3.9165850 &   -4.9045789 &    0.8074595 &    0.2009710 &   -0.0375203 \\
                  &3 & -108.7948038 &  146.4687943 &  -32.9218404 &   -2.5469339 &    0.7759394 \\
\hline                                                                                          \\
                  &1 &   -0.3400174 &    0.7802154 &   -0.3437893 &    0.0824056 &   -0.0066930 \\
H$\delta_{\rm A}$ &2 &   36.1459627 & -178.0526116 &  166.3449941 &  -45.4458738 &    3.7573350 \\
                  &3 &-1381.3144000 & 5283.2867289 &-4334.0624242 & 1125.3443000 &  -90.9465139 \\
\hline                                                                                          \\
                  &1 &   -0.7586478 &    1.6470466 &   -0.6800223 &    0.1306246 &   -0.0086650 \\
H$\gamma_{\rm A}$ &2 &   58.8625436 & -210.0639058 &  174.8989489 &  -45.3356756 &    3.6089852 \\
                  &3 &-2043.5700337 & 6433.0522877 &-4883.0699797 & 1207.4959655 &  -93.9809592 \\
\hline                                                                                          \\
                  &1 &   -0.3060721 &    0.7125074 &   -0.3205038 &    0.0683519 &   -0.0050475 \\
H$\delta_{\rm F}$ &2 &   30.6677991 & -121.8747783 &  101.4148311 &  -26.5930774 &    2.1537733 \\
                  &3 &-1053.7758076 & 3611.9972771 &-2723.9526032 &  682.2344274 &  -54.0667213 \\
\hline                                                                                          \\
                  &1 &   -0.4440136 &    0.9835898 &   -0.4158944 &    0.0801404 &   -0.0053903 \\
H$\gamma_{\rm F}$ &2 &   35.2952498 & -120.5372324 &   95.1523223 &  -24.3827679 &    1.9426707 \\
                  &3 &-1145.3060328 & 3586.2350776 &-2658.3958905 &  655.6380015 &  -51.2487311 \\
\noalign{\smallskip}

\noalign{\smallskip}\hline
\end{tabular}\end{center}
\end{table}

\begin{table}[b]
\caption[]{Similar to Table 1, but for stellar populations older
than 4\,Gyr (Age $\geq$ 4\,Gyr).} \label{Tab:4}
\begin{center}\begin{tabular}{lcrrrrr}
\hline\hline\noalign{\smallskip}%\scriptsize
Index&$j$ &${\rm C_{1j}}$ &${\rm C_{2j}}$ &${\rm C_{3j}}$ &${\rm C_{4j}}$ &${\rm C_{5j}}$\\
\hline
                  &1 &    0.0040188 &   -0.0087338 &    0.0012365 &   -0.0000628 &    0.0000011 \\
CN$_{\rm 1}$      &2 &    0.0699149 &   -0.5550663 &    0.1605598 &   -0.0163788 &    0.0005280 \\
                  &3 &   -4.2835098 &   11.3189732 &   -2.8137761 &    0.2856955 &   -0.0093980 \\
\hline                                                                                          \\
                  &1 &    0.0031607 &   -0.0068581 &    0.0009171 &   -0.0000386 &    0.0000004 \\
CN$_{\rm 2}$      &2 &    0.1358253 &   -0.6142418 &    0.1616565 &   -0.0161493 &    0.0005224 \\
                  &3 &   -5.8469187 &   12.5102930 &   -2.9058639 &    0.2889257 &   -0.0095349 \\
\hline                                                                                          \\
                  &1 &    0.0024019 &   -0.0096646 &    0.0028396 &   -0.0003487 &    0.0000134 \\
Ca4227            &2 &    0.1289259 &   -0.3224843 &   -0.1484763 &    0.0361270 &   -0.0016691 \\
                  &3 &   30.5865753 &  -70.1598679 &   19.3765757 &   -2.1609623 &    0.0788180 \\
\hline                                                                                          \\
                  &1 &    0.1365771 &   -0.4119753 &    0.0710301 &   -0.0055584 &    0.0001697 \\
G4300             &2 &   -3.6369924 &    2.4500207 &    0.5456707 &   -0.0466573 &   -0.0003761 \\
                  &3 &   28.2611775 &  -85.3355224 &   18.7904056 &   -2.1578069 &    0.1062849 \\
\hline                                                                                          \\
                  &1 &    0.0695695 &   -0.1937149 &    0.0295201 &   -0.0017795 &    0.0000416 \\
Fe4383            &2 &    1.5938861 &   -9.3601744 &    2.3994053 &   -0.2263281 &    0.0068871 \\
                  &3 &  -79.2726970 &  189.8780408 &  -38.5276774 &    3.5614204 &   -0.1108842 \\
\hline                                                                                          \\
                  &1 &    0.0113051 &   -0.0260901 &    0.0031068 &   -0.0001249 &    0.0000015 \\
Ca4455            &2 &    0.1825335 &   -2.1257283 &    0.6162648 &   -0.0626674 &    0.0020383 \\
                  &3 &  -18.3660159 &   52.8898164 &  -13.1167084 &    1.3350018 &   -0.0447127 \\
\hline                                                                                          \\
                  &1 &    0.0241263 &   -0.1055130 &    0.0195680 &   -0.0014870 &    0.0000418 \\
Fe4531            &2 &   -0.8767742 &   -0.8527878 &    0.3649493 &   -0.0354193 &    0.0009578 \\
                  &3 &   25.2479921 &  -36.0000062 &    6.6317767 &   -0.5259928 &    0.0185966 \\
\hline                                                                                          \\
                  &1 &    0.0179749 &   -0.0676067 &    0.0083280 &   -0.0002240 &   -0.0000019 \\
Fe4668            &2 &    0.2262582 &   -4.4121200 &    1.5353953 &   -0.1944241 &    0.0071387 \\
                  &3 &  -79.5135908 &  227.1468134 &  -55.0061793 &    6.0957633 &   -0.2199873 \\
\hline                                                                                          \\
                  &1 &   -0.0625565 &    0.2708823 &   -0.0566269 &    0.0045862 &   -0.0001317 \\
H$_\beta$         &2 &    2.8700663 &   -5.7104589 &    0.9274182 &   -0.0705995 &    0.0022889 \\
                  &3 &  -40.1033590 &   83.2554372 &  -18.0609224 &    1.5794001 &   -0.0541752 \\
\hline                                                                                          \\
                  &1 &    0.0326274 &   -0.1721448 &    0.0331789 &   -0.0025354 &    0.0000699 \\
Fe5015            &2 &   -1.1476034 &   -1.5307235 &    0.5946861 &   -0.0674897 &    0.0022440 \\
                  &3 &    7.2743924 &   22.9005565 &   -3.9738686 &    0.4451390 &   -0.0154651 \\
\hline                                                                                          \\
                  &1 &    0.0002583 &   -0.0028680 &    0.0005454 &   -0.0000406 &    0.0000011 \\
Mg$_{\rm 1}$      &2 &    0.0303274 &   -0.1102308 &    0.0115624 &   -0.0003957 &   -0.0000009 \\
                  &3 &   -0.0290541 &    0.0201172 &    0.2573529 &   -0.0302840 &    0.0011283 \\
\hline                                                                                          \\
                  &1 &    0.0003494 &   -0.0047501 &    0.0010250 &   -0.0000857 &    0.0000026 \\
Mg$_{\rm 2}$      &2 &    0.0392009 &   -0.1979893 &    0.0208337 &   -0.0000621 &   -0.0000425 \\
                  &3 &    0.4638829 &   -0.5123346 &    0.6866533 &   -0.0971528 &    0.0039890 \\
\hline                                                                                          \\
                  &1 &   -0.0037660 &   -0.0157034 &    0.0047387 &   -0.0004621 &    0.0000160 \\
Mg$_{\rm b}$      &2 &    0.9698109 &   -4.0019894 &    0.5949929 &   -0.0254183 &    0.0000838 \\
                  &3 &  -10.4139848 &   42.6783675 &   -1.0978610 &   -0.5308182 &    0.0325709 \\
\hline                                                                                          \\
                  &1 &    0.0135064 &   -0.0676152 &    0.0131365 &   -0.0009717 &    0.0000261 \\
Fe5270            &2 &   -0.3787006 &   -1.1509906 &    0.2995794 &   -0.0297612 &    0.0009629 \\
                  &3 &    2.9257542 &   20.0839626 &   -2.9644946 &    0.2873734 &   -0.0103891 \\
\noalign{\smallskip}

\noalign{\smallskip}\hline
\end{tabular}\end{center}
\end{table}

\addtocounter{table}{-1}
\begin{table}[h]
\centering \caption[]{--continued.} \label{Tab:4}
\begin{center}\begin{tabular}{lcrrrrr}
\hline\hline\noalign{\smallskip}%\scriptsize
Index&$j$ &${\rm C_{1j}}$ &${\rm C_{2j}}$ &${\rm C_{3j}}$ &${\rm C_{4j}}$ &${\rm C_{5j}}$\\
\hline                                                                                        \\
                  &1 &    0.0100735 &   -0.0513031 &    0.0112253 &   -0.0009878 &    0.0000303 \\
Fe5335            &2 &   -0.3764122 &   -1.1131799 &   -0.1092792 &    0.0392520 &   -0.0018016 \\
                  &3 &   46.0187216 &  -72.8319926 &   27.5059454 &   -3.2158884 &    0.1133870 \\
\hline                                                                                          \\
                  &1 &    0.0061912 &   -0.0466338 &    0.0093712 &   -0.0007327 &    0.0000206 \\
Fe5406            &2 &    0.0934871 &   -1.3125122 &    0.1939528 &   -0.0108703 &    0.0001877 \\
                  &3 &    1.0004186 &   10.2495446 &    1.2408277 &   -0.2790750 &    0.0120963 \\
\hline                                                                                          \\
                  &1 &    0.0084714 &   -0.0287635 &    0.0050787 &   -0.0003500 &    0.0000089 \\
Fe5709            &2 &   -0.4407574 &    0.0360397 &    0.0939359 &   -0.0153071 &    0.0006047 \\
                  &3 &   -0.8112771 &   19.1722212 &   -4.5075237 &    0.4973342 &   -0.0184540 \\
\hline                                                                                          \\
                  &1 &   -0.0002944 &   -0.0149229 &    0.0032162 &   -0.0003063 &    0.0000099 \\
Fe5782            &2 &    0.2419052 &   -0.2990725 &   -0.1171528 &    0.0246488 &   -0.0010394 \\
                  &3 &   20.8520931 &  -52.8185426 &   16.2838692 &   -1.7500469 &    0.0595642 \\
\hline                                                                                          \\
                  &1 &   -0.0022021 &   -0.0380672 &    0.0095990 &   -0.0008472 &    0.0000260 \\
Na$_{\rm D}$      &2 &    0.3464177 &   -0.9169326 &   -0.0149337 &    0.0195706 &   -0.0010391 \\
                  &3 &   14.6152128 &  -38.0493377 &   12.0162242 &   -1.3802231 &    0.0519050 \\
\hline                                                                                          \\
                  &1 &   -0.0004048 &   -0.0010965 &    0.0002912 &   -0.0000264 &    0.0000008 \\
TiO$_{\rm 1}$     &2 &    0.0098709 &   -0.0060943 &   -0.0015886 &    0.0001050 &   -0.0000058 \\
                  &3 &    0.4022334 &   -0.0305906 &   -0.0854966 &    0.0157629 &   -0.0004537 \\
\hline                                                                                          \\
                  &1 &   -0.0006406 &   -0.0026937 &    0.0006705 &   -0.0000589 &    0.0000018 \\
TiO$_{\rm 2}$     &2 &    0.0411593 &    0.0211796 &   -0.0159178 &    0.0015739 &   -0.0000518 \\
                  &3 &   -0.4009774 &   -0.3431377 &    0.1546906 &   -0.0093757 &    0.0003560 \\
\hline                                                                                          \\
                  &1 &   -0.1822723 &    0.4528323 &   -0.0619613 &    0.0036523 &   -0.0000907 \\
H$\delta_{\rm A}$ &2 &   -9.3242935 &   26.8376168 &   -7.6220614 &    0.6779245 &   -0.0184436 \\
                  &3 &  273.5136304 & -384.4491796 &   87.5549531 &   -6.8624029 &    0.1483407 \\
\hline                                                                                          \\
                  &1 &   -0.2347224 &    0.6562770 &   -0.0985331 &    0.0067082 &   -0.0001900 \\
H$\gamma_{\rm A}$ &2 &   -8.3702686 &   24.2585985 &   -7.8246936 &    0.7094328 &   -0.0186995 \\
                  &3 &  392.7692657 & -531.6987235 &  129.8789712 &  -11.2146737 &    0.2783352 \\
\hline                                                                                          \\
                  &1 &   -0.1106862 &    0.2930026 &   -0.0443820 &    0.0029922 &   -0.0000815 \\
H$\delta_{\rm F}$ &2 &   -3.5060331 &   10.7215430 &   -3.5595302 &    0.3279227 &   -0.0089577 \\
                  &3 &  145.3867861 & -162.5243753 &   44.0186086 &   -3.7498376 &    0.0877458 \\
\hline                                                                                          \\
                  &1 &   -0.1284349 &    0.3643152 &   -0.0583081 &    0.0041664 &   -0.0001196 \\
H$\gamma_{\rm F}$ &2 &   -1.3760493 &    7.0593899 &   -2.8690465 &    0.2678504 &   -0.0067945 \\
                  &3 &  129.2354201 & -186.3215690 &   49.9193880 &   -4.3279933 &    0.0997893 \\
\noalign{\smallskip}

\noalign{\smallskip}\hline
\end{tabular}\end{center}
\end{table}

%
% one-column-wide figure(occupies half-width of a page)
%  -- This is an old way of graphics inclusion with psfig.sty
%------------------------------------------------------------ Fig1: lightcurve
\begin{figure}
   \vspace{2mm}
   \begin{center}
   \hspace{3mm}\psfig{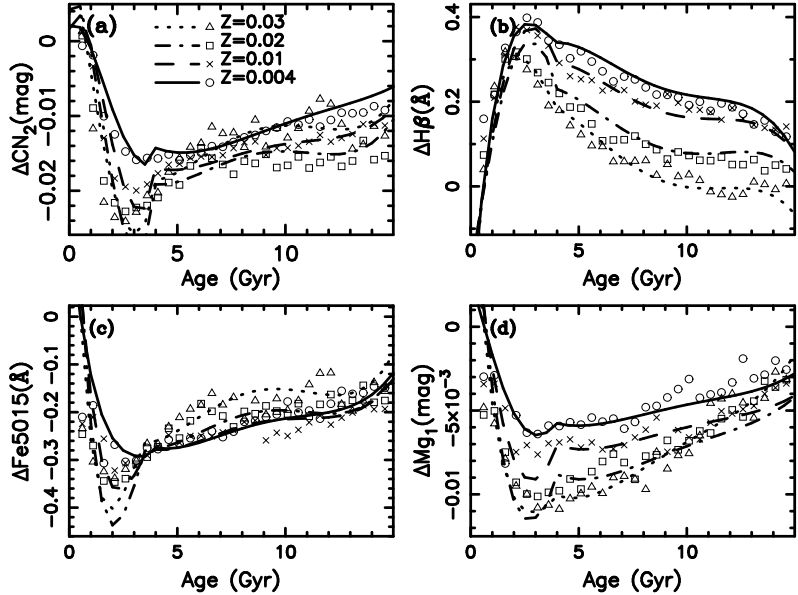}
   \parbox{180mm}{{\vspace{2mm} }}
   \caption{Comparison of fitted and original values for the effects of binary interactions on four Lick indices.
   Circles, crosses, squares, and triangles are for the metallicities of $Z$ = 0.004, 0.01, 0.02, and 0.03, respectively.
   Solid, dashed, dash-dotted, and dotted lines show the fittings for the above four metallicities,
   respectively. The values of y-axes are calculated by subtracting
   the Lick indices of a bsSSP from that of its corresponding (with the same age and metallicity) ssSSP.
   Panels a), b), c), and d) are for CN$_{\rm 2}$, H$\beta$, Fe5015, and Mg$_{\rm 1}$, respectively.}
   \label{Fig:lightcurve-ADAri}
   \end{center}
\end{figure}
%

%
% one-column-wide figure(occupies half-width of a page)
%  -- This is an old way of graphics inclusion with psfig.sty
%------------------------------------------------------------ Fig2: lightcurve
\begin{figure}
   \vspace{2mm}
   \begin{center}
   \hspace{3mm}\psfig{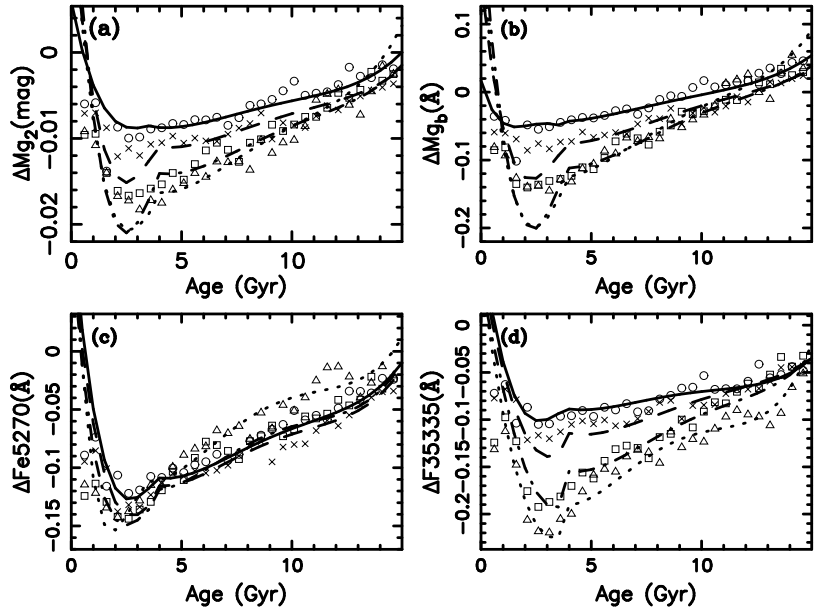}
   \parbox{180mm}{{\vspace{2mm} }}
   \caption{Similar to Fig. 1, but for Mg$_{\rm 2}$, Mg$_{\rm b}$, Fe5270, and Fe5335.}
   \label{Fig:lightcurve-ADAri}
   \end{center}
\end{figure}
%

%
% one-column-wide figure(occupies half-width of a page)
%  -- This is an old way of graphics inclusion with psfig.sty
%------------------------------------------------------------ Fig3: lightcurve
\begin{figure}
   \vspace{2mm}
   \begin{center}
   \hspace{3mm}\psfig{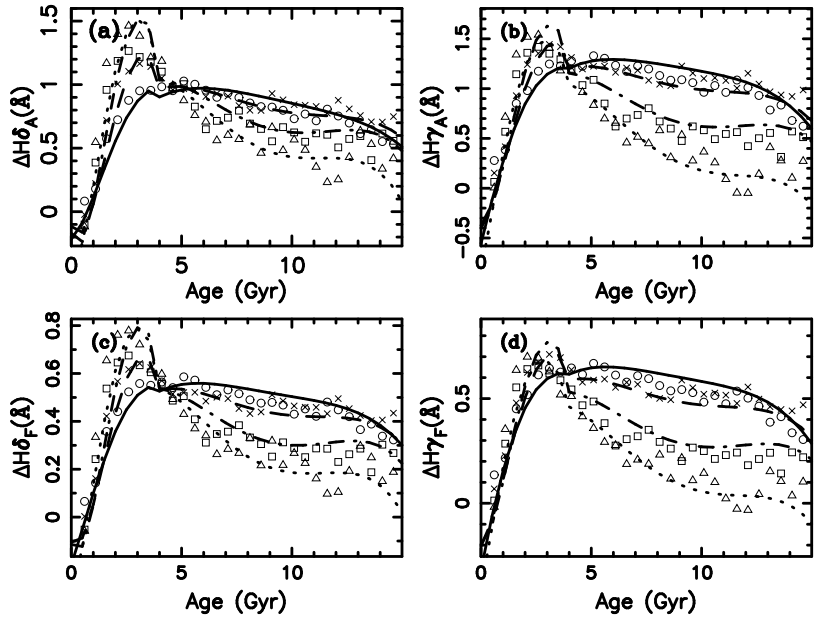}
   \parbox{180mm}{{\vspace{2mm} }}
   \caption{Similar to Fig. 1, but for H$\delta_{\rm A}$, H$\gamma_{\rm A}$, H$\delta_{\rm F}$, and H$\gamma_{\rm F}$.}
   \label{Fig:lightcurve-ADAri}
   \end{center}
\end{figure}

\section{Fitting formulae for effects of binary interactions on 12 colours}
\label{sect:data}
%\hspace{15pt}%                   %% preserved for Editor
Because colours can also be used for stellar population studies, we
fit the formulae for calculating the colour changes cased by binary
interactions when comparing to ssSSPs. One can refer to, e.g.,
\cite{Li:2007potentialcolors, Li:2007effects, Li:2008colourpairs,
Li:2008how}, for the application of colours in stellar population
studies. Some Johnson system colours, Sloan Digital Sky Survey
system (hereafter SDSS-$ugriz$ system) colours, and some composite
colours that consist of both Johnson and an SDSS-$ugriz$ magnitudes
are studied. We only study the colours of populations with $Z \geq$
0.004, because it is difficult to determine the stellar age and
metallicity of metal-poor (e.g., $Z <$ 0.008) populations via
colours under the typical observational uncertainties
(\cite{Li:2008colourpairs}) and metallicity affect the colours of
metal-poor populations stronger. Thus one should use the results
shown here for more metal-poor populations carefully. Because it is
impossible to give the formulae for all colours, we give some ones
for calculating the effects of binary interactions on 12 important
colours, which are sensitive to stellar age or metallicity,
according to the work of \cite{Li:2008colourpairs}. The 12 colours
are $(B-V)$, $(V-K)$, $(I-H)$, $(R-K)$, $(B-K)$, $(I-K)$, $(u-r)$,
$(r-K)$, $(u-R)$, $(u-K)$, $(z-K)$, and $(g-J)$\footnote{Colours
$(r-K)$, $(u-R)$, $(u-K)$, $(z-K)$, and $(g-J)$ are composite
colours. The $UBVRIJHK$ magnitudes are on Johnson system, and
$ugriz$ magnitudes on SDSS-$ugriz$ system.}. Note that $(B-V)$,
$(u-r)$, $(u-R)$, and $(z-K)$ are more sensitive to stellar age and
the others to metallicity. Our work shows that the changes of the
above colours caused by binary interactions can be expressed as
\begin{equation}
    \Delta I' = \sum_{i=1}^{4}
    {{\rm C}_{i}}t^{i-1},
\end{equation}
where $\Delta I'$ is the change of colours caused by binary
interactions, and $t$ is stellar age. The coefficients of the
equation are shown in Table 3. Note that the results for populations
with both Salpeter IMF (standard investigation) and Chabrier IMF are
listed in the table. We can find that equation (2) does not include
the metallicity of populations. The reason is that colours are less
sensitive to metallicity compared to Lick indices and they seem to
be affected by the Mento Carlo method used to generate our star
sample. The fitting of the effects of binary interactions on 12
colours are shown in Figs. 4, 5, and 6. As we see, the fitting
formulae can give average colour changes caused by binary
interactions. However, because the results calculated using equation
(2) have typical errors about 0.02\,mag, some additional
uncertainties may be brought into the results of stellar population
studies.

\begin{table}[]
\caption[]{Coefficients for equation (2). $UBVRIJHKLMN$ magnitudes
are on Johnson system, and $ugriz$ magnitudes are on SDSS-$ugriz$
system.} \label{Tab:4}
\begin{center}\begin{tabular}{c|cccc|cccc}
\hline\hline\noalign{\smallskip}%\scriptsize
IMF    &            &Salpeter     &           &           &            &Chabrier     &           &\\
\hline \multicolumn{9}{c}{Age $<$ 4.2\,Gyr}\\
\hline
Colour &${\rm C_{1}}$ &${\rm C_{2}}$ &${\rm C_{3}}$ &${\rm C_{4}}$ &${\rm C_{1}}$ &${\rm C_{2}}$ &${\rm C_{3}}$ &${\rm C_{4}}$\\
\hline
(B-V)   &-0.014222   &-0.032764    &0.009111   &-0.000722   &-0.020059   &-0.021895    &0.005569   &-0.000449\\
(V-K)   &-0.080134   &-0.093961    &0.038424   &-0.004241   &-0.086556   &-0.062010    &0.021278   &-0.001931\\
(I-H)   &-0.047703   &-0.049909    &0.021701   &-0.002461   &-0.049641   &-0.033034    &0.011729   &-0.001032\\
(R-K)   &-0.043288   &-0.037439    &0.016264   &-0.001812   &-0.042418   &-0.025391    &0.008674   &-0.000701\\
(B-K)   &-0.094328   &-0.126572    &0.047424   &-0.004947   &-0.106087   &-0.084725    &0.027181   &-0.002419\\
(I-K)   &-0.054306   &-0.055956    &0.024500   &-0.002787   &-0.055634   &-0.036844    &0.012924   &-0.001109\\
(u-r)   &-0.056601   &-0.042940    &0.012347   &-0.001002   &-0.063465   &-0.023075    &0.005139   &-0.000410\\
(r-K)   &-0.072748   &-0.083108    &0.035437   &-0.004014   &-0.077569   &-0.054339    &0.019333   &-0.001782\\
(u-R)   &-0.059219   &-0.047988    &0.014264   &-0.001213   &-0.066913   &-0.026108    &0.006208   &-0.000525\\
(u-K)   &-0.129146   &-0.126222    &0.047806   &-0.005016   &-0.140906   &-0.077425    &0.024514   &-0.002201\\
(z-K)   &-0.043288   &-0.037439    &0.016264   &-0.001812   &-0.042418   &-0.025391    &0.008674   &-0.000701\\
(g-J)   &-0.065198   &-0.092019    &0.035014   &-0.003704   &-0.074778   &-0.061406    &0.020278   &-0.001863\\
\hline\noalign{\bigskip}
\hline \multicolumn{9}{c}{Age $\ge$ 4.2\,Gyr}\\
\hline
Colour &${\rm C_{1}}$ &${\rm C_{2}}$ &${\rm C_{3}}$ &${\rm C_{4}}$ &${\rm C_{1}}$ &${\rm C_{2}}$ &${\rm C_{3}}$ &${\rm C_{4}}$\\
\hline
(B-V)   &-0.069795    &0.008858   &-0.000767    &0.000025   &-0.062030    &0.005088   &-0.000384    &0.000014\\
(V-K)   &-0.157537    &0.014833   &-0.001013    &0.000025   &-0.191178    &0.028283   &-0.002893    &0.000103\\
(I-H)   &-0.087919    &0.010592   &-0.000813    &0.000020   &-0.113589    &0.020964   &-0.002196    &0.000076\\
(R-K)   &-0.078362    &0.010635   &-0.000850    &0.000022   &-0.099112    &0.019317   &-0.002018    &0.000069\\
(B-K)   &-0.222114    &0.020910   &-0.001424    &0.000037   &-0.245483    &0.029974   &-0.002870    &0.000102\\
(I-K)   &-0.100169    &0.012567   &-0.000978    &0.000024   &-0.130611    &0.024976   &-0.002622    &0.000090\\
(u-r)   &-0.119987    &0.009885   &-0.001017    &0.000042   &-0.098610    &0.000563   &-0.000163    &0.000019\\
(r-K)   &-0.135641    &0.013834   &-0.001022    &0.000026   &-0.169588    &0.027336   &-0.002877    &0.000102\\
(u-R)   &-0.126930    &0.010203   &-0.001026    &0.000042   &-0.106134    &0.001228   &-0.000236    &0.000022\\
(u-K)   &-0.255286    &0.021748   &-0.001696    &0.000052   &-0.264095    &0.024796   &-0.002561    &0.000101\\
(z-K)   &-0.078362    &0.010635   &-0.000850    &0.000022   &-0.099112    &0.019317   &-0.002018    &0.000069\\
(g-J)   &-0.154783    &0.014291   &-0.000930    &0.000024   &-0.171114    &0.020427   &-0.001897    &0.000067\\
\noalign{\smallskip}\hline
\end{tabular}\end{center}
\end{table}

%
% one-column-wide figure(occupies half-width of a page)
%  -- This is an old way of graphics inclusion with psfig.sty
%------------------------------------------------------------ Fig4: lightcurve
\begin{figure}
   \vspace{2mm}
   \begin{center}
   \hspace{3mm}\psfig{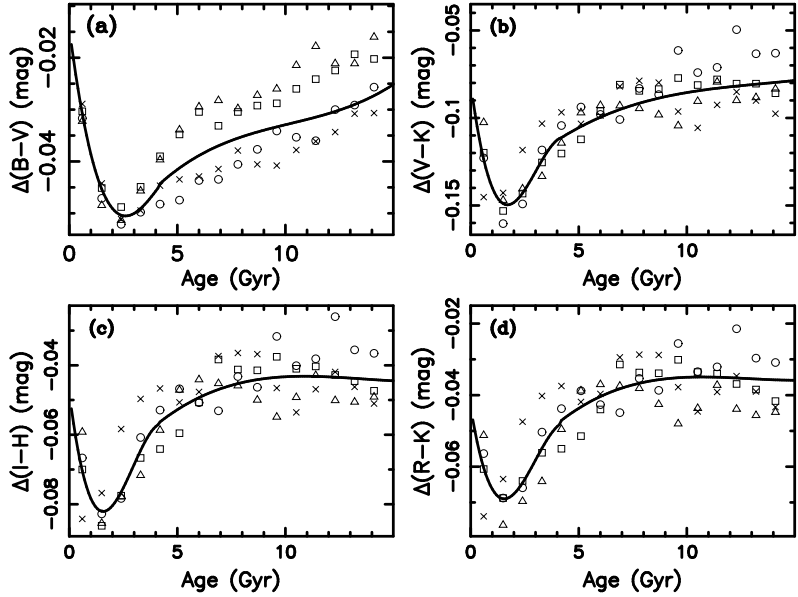}
   \parbox{180mm}{{\vspace{2mm} }}
   \caption{Fittings for the effects of binary interactions on four colours of populations.
   Circles, crosses, squares, and triangles show the values obtained directly from comparing
   the colours of bsSSPs and ssSSPs and are for metallicities of 0.004, 0.01, 0.02, and 0.03, respectively.
   Solid lines show the fittings. The y-axis is obtained by subtracting
   the colour of a bsSSP from that of an ssSSP (with the same age and metallicity as the bsSSP).
   The four panels are for $(B-V)$, $(V-K)$, $(I-H)$, and $(R-K)$, respectively.}
   \label{Fig:lightcurve-ADAri}
   \end{center}
\end{figure}
%

%
% one-column-wide figure(occupies half-width of a page)
%  -- This is an old way of graphics inclusion with psfig.sty
%------------------------------------------------------------ Fig5: lightcurve
\begin{figure}
   \vspace{2mm}
   \begin{center}
   \hspace{3mm}\psfig{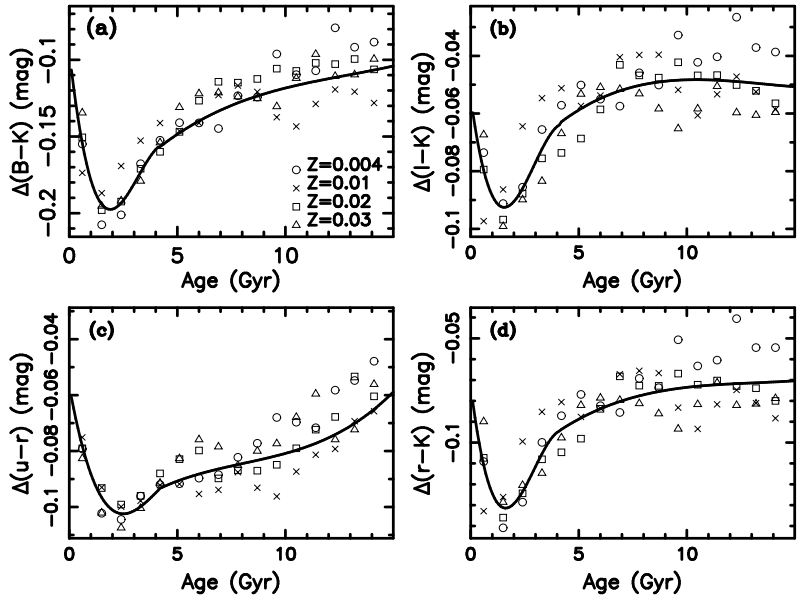}
   \parbox{180mm}{{\vspace{2mm} }}
   \caption{Similar to Fig. 4, but for $(B-K)$, $(I-K)$, $(u-r)$, and $(r-K)$. }
   \label{Fig:lightcurve-ADAri}
   \end{center}
\end{figure}
%

%
% one-column-wide figure(occupies half-width of a page)
%  -- This is an old way of graphics inclusion with psfig.sty
%------------------------------------------------------------ Fig6: lightcurve
\begin{figure}
   \vspace{2mm}
   \begin{center}
   \hspace{3mm}\psfig{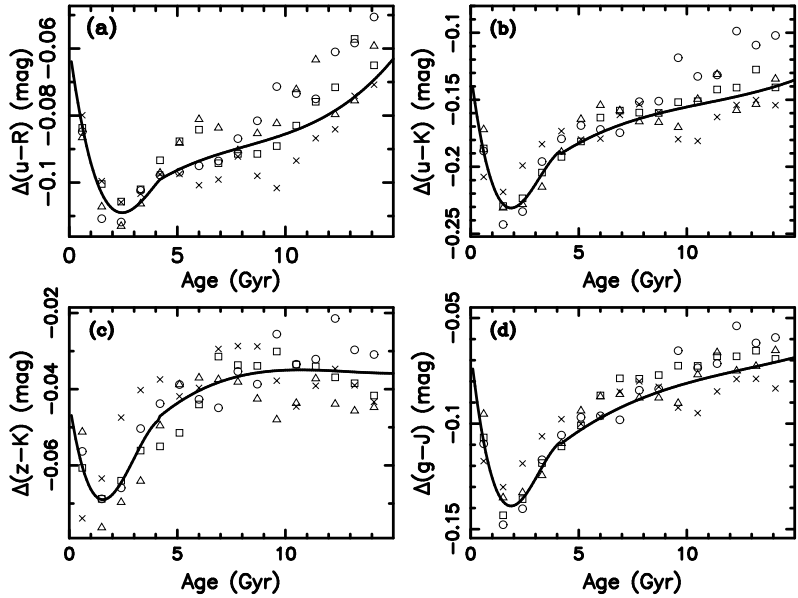}
   \parbox{180mm}{{\vspace{2mm} }}
   \caption{Similar to Fig. 4, but for $(u-R)$, $(u-K)$, $(z-K)$, and $(g-J)$. }
   \label{Fig:lightcurve-ADAri}
   \end{center}
\end{figure}

\section{Discussion and conclusions}
We present some formulae for conveniently computing the changes
caused by binary interactions in 25 Lick indices and 12 colours,
compared to the indices of single star stellar populations (ssSSPs).
It is shown that the fitting formulae presented in the paper can
reproduce the changes in Lick indices caused by binary interactions
with small errors and can be used to estimate similar changes in
colours. It is also found that binary interactions make
age-sensitive Lick indices (not only H$\beta$, but also
$H\delta_{\rm A}$, $H\delta_{\rm F}$, $H\gamma_{\rm A}$,
$H\gamma_{\rm F}$) larger, while making metallicity-sensitive
indices less compared to those of ssSSPs. This is useful for
estimating the effects of binary evolution on the results of stellar
population studies and for adding the effects of binary interactions
into ssSSP models. Therefore, when an age-sensitive Lick index is
used together with a metallicity-sensitive Lick index to determine
the ages and metallicities of populations, younger ages will be
obtained, especially for metal-poor populations, see also
\cite{Li:2008how}. Note that only binary star stellar populations
(bsSSPs) and ssSSPs with four metallicities ($Z$ = 0.004, 0.01,
0.02, and 0.03) are used in the work. This is actually limited by
the metallicity coverage of stellar population model and the less
sensitivities of colours to metallicity. Thus the results are more
suitable for studying metal-rich ($Z \geq$ 0.004) populations,
because the differences between integrated peculiarities of
populations with various metallicities seem larger for metal-poor
populations. In addition, although different formulae are presented
for populations with various initial mass functions (IMFs), the
changes calculated via two kinds of formulae (the formulae for
populations with Salpeter and Chabrier IMFs) are similar. In other
words, the changes calculated by the formulae obtained using
populations with Salpeter IMF or Chabrier IMF can give us some
pictures for the effects of binary interactions. Furthermore,
because the Monte Carlo technique used to generate the binary sample
of stellar populations make the evolution of integrated
peculiarities of populations unsmooth, some results, especially,
those for colours, may be somewhat rough. The additional
uncertainties involved should be taken into account. If possible, we
will give more detailed studies in the future.

\begin{acknowledgements}
We thank the referee, Profs. Gang Zhao, Xu Zhou, Licai Deng, Xu
Kong, Tinggui Wang, and Li Zhang for useful suggestions or
discussions. This work is supported by the Chinese National Science
Foundation (Grant Nos. 10433030, 10521001, 2007CB815406).
\end{acknowledgements}

%\bibliography{tex}
%\bibliographystyle{apj}

\label{lastpage}

%\end{document}
%%==^..^============== the END of cjaa.tex ===================^_^==

\newpage
\appendix
\section{Coefficients for calculating the effects of binary interactions on 25 Lick indices of populations with Chabrier IMF}

\begin{table}[]
\caption[]{Coefficients for equation (1). The coefficients are
obtained via stellar populations with Chabrier IMF and can be used
for populations younger than 3.5\,Gyr (Age $<$ 3.5\,Gyr).}
\label{Tab:4}
\begin{center}\begin{tabular}{lcrrrrr}
\hline\hline\noalign{\smallskip}%\scriptsize
Index&$j$ &${\rm C_{1j}}$ &${\rm C_{2j}}$ &${\rm C_{3j}}$ &${\rm C_{4j}}$ &${\rm C_{5j}}$\\
\hline
                  &1 &    0.0022919 &    0.0011051 &   -0.0063997 &    0.0015981 &   -0.0001080 \\
CN$_{\rm 1}$      &2 &    0.6329579 &   -0.9595824 &    0.2427491 &   -0.0011151 &   -0.0025634 \\
                  &3 &   -4.5940253 &    7.9284009 &   -3.9250539 &    0.6480545 &   -0.0256989 \\
\hline                                                                                          \\
                  &1 &    0.0012774 &    0.0032128 &   -0.0069053 &    0.0016921 &   -0.0001156 \\
CN$_{\rm 2}$      &2 &    0.6334017 &   -1.0656913 &    0.3400438 &   -0.0309074 &   -0.0000476 \\
                  &3 &   -4.4675288 &    9.5105511 &   -5.7521116 &    1.2233685 &   -0.0747440 \\
\hline                                                                                          \\
                  &1 &    0.0795038 &   -0.1571070 &    0.0740613 &   -0.0128663 &    0.0007442 \\
Ca4227            &2 &  -11.9558663 &   20.8851588 &   -9.0611781 &    1.3421325 &   -0.0622930 \\
                  &3 &  342.8235941 & -529.6545151 &  161.4994646 &  -12.9148165 &   -0.1134096 \\
\hline                                                                                          \\
                  &1 &    0.7660019 &   -1.5947618 &    0.6186025 &   -0.0971560 &    0.0052551 \\
G4300             &2 &  -87.0440009 &  155.4728799 &  -65.2201097 &    9.5672170 &   -0.4355927 \\
                  &3 & 2358.1318663 &-4145.5253913 & 1656.4267226 & -221.2634252 &    8.6149062 \\
\hline                                                                                          \\
                  &1 &    0.6089788 &   -0.9690479 &    0.3009590 &   -0.0329757 &    0.0009352 \\
Fe4383            &2 &  -72.6995453 &  111.7196061 &  -45.0101690 &    6.0333694 &   -0.2331824 \\
                  &3 & 2100.1581312 &-3361.5113576 & 1335.7159550 & -174.8630448 &    6.5371492 \\
\hline                                                                                          \\
                  &1 &    0.0681495 &   -0.1369378 &    0.0592712 &   -0.0105410 &    0.0006446 \\
Ca4455            &2 &   -7.3288591 &   13.2396365 &   -7.0194851 &    1.2883776 &   -0.0767611 \\
                  &3 &  258.9163092 & -510.1718798 &  260.9018384 &  -46.2914894 &    2.6884347 \\
\hline                                                                                          \\
                  &1 &    0.2564424 &   -0.5328445 &    0.2229634 &   -0.0364862 &    0.0020371 \\
Fe4531            &2 &  -26.0845865 &   42.1057901 &  -17.3504899 &    2.4029055 &   -0.0987445 \\
                  &3 &  851.0453910 &-1492.6215470 &  616.0271336 &  -86.7449029 &    3.7433351 \\
\hline                                                                                          \\
                  &1 &    0.1275239 &   -0.2791766 &    0.1024538 &   -0.0161675 &    0.0009372 \\
Fe4668            &2 &   -5.2389930 &    7.6979073 &   -2.3867283 &    0.1846930 &   -0.0012464 \\
                  &3 &  169.0223930 & -536.0764913 &  339.6448905 &  -65.7521400 &    4.1760430 \\
\hline                                                                                          \\
                  &1 &   -0.4293708 &    0.8954661 &   -0.3155113 &    0.0432326 &   -0.0020233 \\
H$_\beta$         &2 &   30.1096288 &  -49.0583758 &   16.8580314 &   -1.8838154 &    0.0479103 \\
                  &3 & -767.8393985 & 1229.9316335 & -430.0651511 &   45.5750506 &   -0.8795314 \\
\hline                                                                                          \\
                  &1 &    0.2395560 &   -0.3782478 &    0.1010824 &   -0.0116568 &    0.0005056 \\
Fe5015            &2 &   -3.3348359 &  -34.6665677 &   26.0424965 &   -5.8877139 &    0.4132806 \\
                  &3 &  413.3734390 &   45.5824955 & -217.1527754 &   66.7394428 &   -5.4107732 \\
\hline                                                                                          \\
                  &1 &    0.0057639 &   -0.0122970 &    0.0048864 &   -0.0007787 &    0.0000435 \\
Mg$_{\rm 1}$      &2 &   -0.3355850 &    0.6283524 &   -0.2850889 &    0.0319693 &   -0.0005680 \\
                  &3 &   13.7239823 &  -27.6095038 &   10.7666288 &   -1.0488923 &    0.0060013 \\
\hline                                                                                          \\
                  &1 &    0.0123326 &   -0.0261803 &    0.0111134 &   -0.0018290 &    0.0001038 \\
Mg$_{\rm 2}$      &2 &   -0.7115473 &    1.0153443 &   -0.3612812 &    0.0205355 &    0.0016030 \\
                  &3 &   27.2273352 &  -44.1981147 &   13.4974485 &   -0.5084821 &   -0.0847616 \\
\hline                                                                                          \\
                  &1 &    0.1259541 &   -0.2256908 &    0.0880777 &   -0.0124308 &    0.0005828 \\
Mg$_{\rm b}$      &2 &   -8.2670312 &    5.7969056 &    0.6635323 &   -0.9033180 &    0.1008251 \\
                  &3 &  311.7946049 & -348.4131106 &   36.5555001 &   20.8760850 &   -2.8507846 \\
\hline                                                                                          \\
                  &1 &    0.1526957 &   -0.3633561 &    0.1689542 &   -0.0304730 &    0.0018686 \\
Fe5270            &2 &  -16.8516642 &   33.9221551 &  -18.3927017 &    3.5096217 &   -0.2189850 \\
                  &3 &  587.4969012 &-1243.1886864 &  678.5522252 & -131.0104877 &    8.3026556 \\
\noalign{\smallskip}

\noalign{\smallskip}\hline
\end{tabular}\end{center}
\end{table}

\addtocounter{table}{-1}
\begin{table}[]
\centering \caption[]{--continued.} \label{Tab:4}
\begin{center}\begin{tabular}{lcrrrrr}
\hline\hline\noalign{\smallskip}%\scriptsize
Index&$j$ &${\rm C_{1j}}$ &${\rm C_{2j}}$ &${\rm C_{3j}}$ &${\rm C_{4j}}$ &${\rm C_{5j}}$\\
\hline                                                                                        \\
                  &1 &    0.2076133 &   -0.4791791 &    0.2222679 &   -0.0377229 &    0.0021281 \\
Fe5335            &2 &  -32.4556974 &   65.1973802 &  -31.8891247 &    5.4057781 &   -0.2980594 \\
                  &3 &  935.2591701 &-1881.6098576 &  872.5249261 & -142.2695305 &    7.5855211 \\
\hline                                                                                          \\
                  &1 &    0.0953714 &   -0.2161430 &    0.0970755 &   -0.0171741 &    0.0010419 \\
Fe5406            &2 &   -3.9285567 &    6.4393293 &   -3.6907976 &    0.6355834 &   -0.0343946 \\
                  &3 &  174.8445638 & -345.4890330 &  172.7005100 &  -28.2035654 &    1.4789489 \\
\hline                                                                                          \\
                  &1 &    0.0665533 &   -0.1333817 &    0.0460766 &   -0.0056987 &    0.0002133 \\
Fe5709            &2 &   -8.4184254 &   11.6155577 &   -2.8870447 &    0.0168183 &    0.0277514 \\
                  &3 &  229.4380018 & -346.7249858 &  109.9625205 &   -7.5674339 &   -0.2639790 \\
\hline                                                                                          \\
                  &1 &    0.0724796 &   -0.1553226 &    0.0682675 &   -0.0111777 &    0.0006131 \\
Fe5782            &2 &  -12.3652716 &   23.1281644 &  -10.7411540 &    1.7511767 &   -0.0935373 \\
                  &3 &  379.6596477 & -665.9564600 &  268.9228630 &  -38.4448161 &    1.7703055 \\
\hline                                                                                          \\
                  &1 &    0.0714688 &   -0.1651977 &    0.0664805 &   -0.0103341 &    0.0005679 \\
Na$_{\rm D}$      &2 &   -1.5605878 &   -4.9385375 &    5.3394247 &   -1.5674185 &    0.1304109 \\
                  &3 &  148.3896060 &  -24.0899505 & -130.0698652 &   49.4638805 &   -4.4865514 \\
\hline                                                                                          \\
                  &1 &   -0.0115706 &    0.0112891 &   -0.0010864 &   -0.0006067 &    0.0000840 \\
TiO$_{\rm 1}$     &2 &    2.0705973 &   -2.9503555 &    0.8324854 &   -0.0401368 &   -0.0040348 \\
                  &3 &  -44.5370053 &   67.5680594 &  -20.7546877 &    1.4636658 &    0.0462995 \\
\hline                                                                                          \\
                  &1 &   -0.0125496 &    0.0097243 &   -0.0001968 &   -0.0008541 &    0.0001063 \\
TiO$_{\rm 2}$     &2 &    2.4479466 &   -3.6429429 &    1.1610628 &   -0.0929928 &   -0.0014161 \\
                  &3 &  -49.2619296 &   79.6645168 &  -27.3299096 &    2.6274999 &   -0.0183345 \\
\hline                                                                                          \\
                  &1 &   -0.1701616 &    0.1253346 &    0.2496936 &   -0.0688490 &    0.0048300 \\
H$\delta_{\rm A}$ &2 &  -22.7201592 &   34.6318625 &  -16.9398410 &    2.8138566 &   -0.1414102 \\
                  &3 &   35.0832106 & -330.3012402 &  581.7572135 & -167.8682957 &   12.8850332 \\
\hline                                                                                          \\
                  &1 &   -1.5856131 &    2.7085039 &   -0.9113125 &    0.1227389 &   -0.0055250 \\
H$\gamma_{\rm A}$ &2 &  180.2526812 & -282.6477080 &  118.3329534 &  -17.2376118 &    0.7715612 \\
                  &3 &-5010.4626342 & 7614.1515475 &-2881.3416065 &  351.0737254 &  -11.0363129 \\
\hline                                                                                          \\
                  &1 &   -0.1251870 &    0.1470139 &    0.1086295 &   -0.0343830 &    0.0025515 \\
H$\delta_{\rm F}$ &2 &  -17.3098016 &   30.7053467 &  -17.0765539 &    3.2519508 &   -0.1978084 \\
                  &3 &  137.3583135 & -435.8552796 &  470.4948769 & -124.2942197 &    9.3212486 \\
\hline                                                                                          \\
                  &1 &   -0.6254725 &    1.1213832 &   -0.3612095 &    0.0484349 &   -0.0022475 \\
H$\gamma_{\rm F}$ &2 &   53.4441919 &  -86.7513473 &   36.9363160 &   -5.4971981 &    0.2510974 \\
                  &3 &-1532.4775410 & 2361.4227142 & -884.2807841 &  102.2525869 &   -2.6516141 \\
\noalign{\smallskip}\hline
\end{tabular}\end{center}
\end{table}

\begin{table}[]
\caption[]{Similar to Table A.1, but for stellar populations older
than 3.5\,Gyr (Age $\geq$ 3.5\,Gyr).} \label{Tab:4}
\begin{center}\begin{tabular}{lcrrrrr}
\hline\hline\noalign{\smallskip}%\scriptsize
Index&$j$ &${\rm C_{1j}}$ &${\rm C_{2j}}$ &${\rm C_{3j}}$ &${\rm C_{4j}}$ &${\rm C_{5j}}$\\
\hline
                  &1 &    0.0050070 &   -0.0101123 &    0.0016866 &   -0.0001101 &    0.0000026 \\
CN$_{\rm 1}$      &2 &   -0.7387795 &   -0.0268996 &    0.0467357 &   -0.0067710 &    0.0002559 \\
                  &3 &    1.8158227 &    8.1027523 &   -2.3010584 &    0.2598589 &   -0.0092777 \\
\hline                                                                                          \\
                  &1 &    0.0048267 &   -0.0086639 &    0.0014493 &   -0.0000926 &    0.0000021 \\
CN$_{\rm 2}$      &2 &   -0.6713808 &   -0.0767864 &    0.0454875 &   -0.0063150 &    0.0002435 \\
                  &3 &    2.4347829 &    7.4817292 &   -1.9820797 &    0.2264804 &   -0.0083295 \\
\hline                                                                                          \\
                  &1 &    0.0253654 &   -0.0209602 &    0.0048307 &   -0.0004936 &    0.0000168 \\
Ca4227            &2 &   -3.5901788 &    1.5319737 &   -0.4799447 &    0.0598457 &   -0.0022187 \\
                  &3 &  108.6370369 & -107.2440863 &   25.4608949 &   -2.5477372 &    0.0859891 \\
\hline                                                                                          \\
                  &1 &    0.2092102 &   -0.4271461 &    0.0709524 &   -0.0053369 &    0.0001564 \\
G4300             &2 &  -32.0561313 &   13.8970486 &   -0.9523866 &    0.0227085 &   -0.0010081 \\
                  &3 &  388.0491119 & -107.1700035 &   -9.9303420 &    2.4271319 &   -0.0776931 \\
\hline                                                                                          \\
                  &1 &    0.0852039 &   -0.2016152 &    0.0323234 &   -0.0021218 &    0.0000535 \\
Fe4383            &2 &  -12.6431474 &   -2.2748662 &    1.1500142 &   -0.1342407 &    0.0045588 \\
                  &3 &   60.8341421 &  179.0155270 &  -49.0631558 &    5.2647148 &   -0.1806644 \\
\hline                                                                                          \\
                  &1 &    0.0439274 &   -0.0430992 &    0.0064975 &   -0.0004070 &    0.0000095 \\
Ca4455            &2 &   -6.6685854 &    1.4457768 &   -0.0535448 &   -0.0104817 &    0.0006266 \\
                  &3 &  114.6093932 &   -5.2478791 &   -4.3355245 &    0.8004669 &   -0.0341767 \\
\hline                                                                                          \\
                  &1 &    0.0586416 &   -0.1084756 &    0.0185745 &   -0.0013543 &    0.0000372 \\
Fe4531            &2 &  -16.5223730 &    5.9250743 &   -0.6822331 &    0.0345448 &   -0.0007249 \\
                  &3 &  327.8776772 & -131.2118033 &   13.2546833 &   -0.3619352 &   -0.0008992 \\
\hline                                                                                          \\
                  &1 &    0.0579868 &   -0.0883225 &    0.0125230 &   -0.0005465 &    0.0000058 \\
Fe4668            &2 &  -14.2549909 &    3.6173449 &   -0.0066150 &   -0.0775483 &    0.0041877 \\
                  &3 &  256.6237954 &   70.1814681 &  -29.9674944 &    4.6116691 &   -0.1954322 \\
\hline                                                                                          \\
                  &1 &    0.1188078 &    0.1462153 &   -0.0301692 &    0.0023434 &   -0.0000653 \\
H$_\beta$         &2 &    5.2527398 &   -5.1845420 &    0.5595370 &   -0.0222041 &    0.0004090 \\
                  &3 &   -6.8462900 &   -4.4966224 &    9.4442175 &   -1.3358200 &    0.0457839 \\
\hline                                                                                          \\
                  &1 &    0.0327476 &   -0.1444852 &    0.0252383 &   -0.0018120 &    0.0000483 \\
Fe5015            &2 &  -22.2627396 &    7.4518761 &   -0.7875696 &    0.0244834 &    0.0000504 \\
                  &3 &  369.7815953 &  -72.0823837 &   -1.2940653 &    1.1387862 &   -0.0545511 \\
\hline                                                                                          \\
                  &1 &   -0.0029387 &   -0.0005618 &    0.0000571 &   -0.0000025 &    0.0000001 \\
Mg$_{\rm 1}$      &2 &    0.0681456 &   -0.2239214 &    0.0461259 &   -0.0037007 &    0.0001017 \\
                  &3 &   -1.7505218 &    4.5445049 &   -1.1536134 &    0.1112986 &   -0.0034731 \\
\hline                                                                                          \\
                  &1 &   -0.0022995 &   -0.0025539 &    0.0005160 &   -0.0000415 &    0.0000013 \\
Mg$_{\rm 2}$      &2 &   -0.3140984 &   -0.1102010 &    0.0232351 &   -0.0016550 &    0.0000414 \\
                  &3 &    3.8491416 &    2.2473119 &   -0.6944256 &    0.0739731 &   -0.0024430 \\
\hline                                                                                          \\
                  &1 &    0.0165715 &   -0.0215425 &    0.0043674 &   -0.0002914 &    0.0000068 \\
Mg$_{\rm b}$      &2 &   -8.0187674 &    0.1612416 &    0.0564198 &   -0.0091935 &    0.0004596 \\
                  &3 &  158.6306124 &  -15.8632521 &    0.7547564 &    0.1990004 &   -0.0142351 \\
\hline                                                                                          \\
                  &1 &   -0.0053182 &   -0.0449585 &    0.0074887 &   -0.0004761 &    0.0000116 \\
Fe5270            &2 &   -6.2888460 &    0.8351887 &    0.0980695 &   -0.0229461 &    0.0009453 \\
                  &3 &  106.4680655 &   11.8581491 &   -9.0607405 &    1.2154258 &   -0.0455243 \\
\noalign{\smallskip}

\noalign{\smallskip}\hline
\end{tabular}\end{center}
\end{table}

\addtocounter{table}{-1}
\begin{table}[]
\centering \caption[]{--continued.} \label{Tab:4}
\begin{center}\begin{tabular}{lcrrrrr}
\hline\hline\noalign{\smallskip}%\scriptsize
Index&$j$ &${\rm C_{1j}}$ &${\rm C_{2j}}$ &${\rm C_{3j}}$ &${\rm C_{4j}}$ &${\rm C_{5j}}$\\
                  &1 &   -0.0299036 &   -0.0168634 &    0.0034824 &   -0.0003504 &    0.0000128 \\
Fe5335            &2 &   -0.8764299 &   -2.3631112 &    0.3216860 &   -0.0034947 &   -0.0004859 \\
                  &3 &  -19.1581245 &   20.3079448 &    1.5735482 &   -0.7362636 &    0.0365434 \\
\hline                                                                                          \\
                  &1 &   -0.0273466 &   -0.0191557 &    0.0032597 &   -0.0002321 &    0.0000069 \\
Fe5406            &2 &   -1.6859288 &   -1.4089294 &    0.3578634 &   -0.0309966 &    0.0008859 \\
                  &3 &    3.2750120 &   49.7275784 &  -12.4265243 &    1.1493382 &   -0.0351427 \\
\hline                                                                                          \\
                  &1 &    0.0058300 &   -0.0239914 &    0.0038273 &   -0.0002377 &    0.0000055 \\
Fe5709            &2 &   -3.1695351 &    1.3275974 &   -0.1282929 &    0.0015248 &    0.0001363 \\
                  &3 &   62.3172407 &   -5.6986762 &   -1.2147039 &    0.3086961 &   -0.0142722 \\
\hline                                                                                          \\
                  &1 &   -0.0351759 &    0.0072524 &   -0.0009035 &   -0.0000295 &    0.0000040 \\
Fe5782            &2 &    2.9123930 &   -2.1632509 &    0.2392269 &    0.0009764 &   -0.0005652 \\
                  &3 &  -87.2972968 &   23.4730212 &    0.8426133 &   -0.6095760 &    0.0321487 \\
\hline                                                                                          \\
                  &1 &   -0.0349510 &   -0.0109978 &    0.0032700 &   -0.0003007 &    0.0000101 \\
Na$_{\rm D}$      &2 &   -3.1483417 &   -0.0799733 &    0.0306158 &    0.0017984 &   -0.0001874 \\
                  &3 &   74.8532596 &  -28.0695426 &    1.8470812 &    0.0045464 &   -0.0013129 \\
\hline                                                                                          \\
                  &1 &   -0.0019580 &   -0.0001078 &    0.0001071 &   -0.0000133 &    0.0000005 \\
TiO$_{\rm 1}$     &2 &   -0.0663381 &    0.0341998 &   -0.0108762 &    0.0008305 &   -0.0000198 \\
                  &3 &    3.2582308 &   -1.5575893 &    0.2498990 &   -0.0085626 &   -0.0000514 \\
\hline                                                                                          \\
                  &1 &   -0.0055418 &    0.0002662 &    0.0000839 &   -0.0000138 &    0.0000006 \\
TiO$_{\rm 2}$     &2 &    0.1861284 &   -0.0836849 &    0.0091382 &   -0.0007762 &    0.0000260 \\
                  &3 &   -4.2466215 &    3.0328182 &   -0.7327127 &    0.0800970 &   -0.0027508 \\
\hline                                                                                          \\
                  &1 &   -0.4834854 &    0.6747627 &   -0.1136504 &    0.0081802 &   -0.0002208 \\
H$\delta_{\rm A}$ &2 &   48.2162849 &   -8.6259083 &   -0.5680772 &    0.1330894 &   -0.0044290 \\
                  &3 & -391.8235480 &  -44.3395429 &   41.4350353 &   -5.4239881 &    0.1828454 \\
\hline                                                                                          \\
                  &1 &   -0.5340026 &    0.8499248 &   -0.1407715 &    0.0102053 &   -0.0002837 \\
H$\gamma_{\rm A}$ &2 &   57.5218630 &  -12.4432398 &   -1.0323349 &    0.2184516 &   -0.0069865 \\
                  &3 & -381.7703663 & -237.3724815 &  109.5206537 &  -12.9300524 &    0.4258356 \\
\hline                                                                                          \\
                  &1 &   -0.2328216 &    0.3790464 &   -0.0645870 &    0.0047630 &   -0.0001318 \\
H$\delta_{\rm F}$ &2 &   28.4618861 &   -8.4343118 &    0.2882612 &    0.0224536 &   -0.0007588 \\
                  &3 & -256.8427724 &   41.7473144 &   12.0094480 &   -2.0067178 &    0.0654486 \\
\hline                                                                                          \\
                  &1 &   -0.2081433 &    0.4185126 &   -0.0709583 &    0.0052155 &   -0.0001461 \\
H$\gamma_{\rm F}$ &2 &   26.6976093 &   -8.4741360 &    0.0074237 &    0.0631139 &   -0.0020772 \\
                  &3 & -176.6172339 &  -82.1781969 &   45.6384532 &   -5.4871418 &    0.1779060 \\
\noalign{\smallskip}

\noalign{\smallskip}\hline
\end{tabular}\end{center}
\end{table}

\end{document}